\documentstyle[prl,aps,epsf,epsfig,array,eqsecnum]{revtex}

\newcommand{\be}{\begin{equation}}
\newcommand{\ee}{\end{equation}}
\newcommand{\ben}{\begin{eqnarray}}
\newcommand{\een}{\end{eqnarray}}

\begin{document}
\setlength{\topmargin}{-2ex}

\title{Gauge fluctuations and transition temperature for
superconducting wires}

\author{A.P.C. Malbouisson, Y.W. Milla, I. Roditi}
\address{CBPF/MCT, Rua Dr. Xavier Sigaud, 150, Rio de Janeiro RJ, Brazil}

\date{today}
\maketitle
\begin{abstract}

We consider  the Ginzburg-Landau model, confined in an infinitely
long rectangular wire of cross-section $L_{1}\times L_{2}$. Our
approach is based on the Gaussian effective potential in the
transverse unitarity gauge, which allows to treat  gauge
contributions in a compact form. The contributions from the scalar
self-interaction and from the gauge fluctuations are clearly
identified. Using techniques from dimensional and $zeta$-function
regularizations, modified by the confinement conditions, we
investigate the critical temperature for a wire of
transverse dimensions $L_1$, $L_2$.  Taking the mass term in
 the form $m_{0}^2=a(T/T_0 - 1)$, where $T_0$ is the bulk transition
temperature,
we obtain equations for the critical temperature as a function
of the $L_{i}'s$  and of $T_{0}$, and determine the limiting sizes
sustaining the transition.
 A qualitative comparison with some experimental observations is done.
\\

\noindent PACS number(s): 11.10.Jj; 11.10.Kk; 11.15.Pg\\
\\

\end{abstract}

\section{Introduction}

In the present work we discuss the critical behavior of the
Ginzburg-Landau model compactified in
   spatial dimensions and we implement the gauge contributions
using the Gaussian effective potential formalism. Spontaneous
symmetry breaking is obtained by taking the bare mass coefficient
in the Hamiltonian parametrized as $m_{0}^2=a(T/T_0 - 1)$,
with $a\,> 0$ and the parameter $T$ varying in an interval
containing the bulk transition
temperature $T_0$. With this choice, considering the system
confined in an infinitely long rectangular cylinder with cross-section
area $A=L_{1}\times L_{2}$,
 we obtain the Ginzburg-Landau model
describing phase
transitions in samples of a material in the form  a wire.
  This generalizes to wires
recent results obtained for materials in the form of films
\cite{isaque}. We investigate the behavior of the system taking
into account gauge fluctuations, which means that charged
transitions are included in our work. We are particularly
interested in the problem of how the critical behavior depends on
the dimensions $L_{1},L_{2}$. This study is done by means of the
Gaussian Effective Potential (GEP) as developed in Refs.
\cite{Gep1,Gep2,Gep3,Gep4}, together with a spatial
compactification mechanism introduced in  recent
publications\cite{Ademir,JMario}.

A central ingredient in our approach relies on the topological nature of the Matsubara
imaginary-time formalism. In order to perform the calculation of the
partition function in a quantum field theory, the Matsubara prescription results to
be equivalent to a path-integral calculated on \ $R^{D-1}\times S^{1}$,
where $S^{1}$ is a circle of circumference $\beta =1/T.$ This
result was demonstrated at the one-loop level in Ref. \cite{pol1} and
has been assumed to be valid for higher orders \cite{pol2}.
Relying on the fact that for Euclidian field theories inverse imaginary time and
spatial coordinates are on the same footing,
such a topological result has  been generalized to treat
different physical situations in which fields are confined in higher purely {\it spatial}
dimensions, considering the Matsubara mechanism on a $R^{D-d}\times
S^{1_{1}}\times S^{1_{2}}...\times S^{1_{d}}$ topology, describing
  space confinement in a $d$-dimensional subspace \cite{Ademir,JMario,GNN}.
As a consequence the Matsubara formalism can be thought,
in a generalized way, as a mechanism to deal  with
spatial constraints  in a field theory model. In this situation, for
consistency, the fields fulfill periodic
(antiperiodic) boundary conditions for bosons (fermions).
In any case we infer from the above discussion that we
are justified to consider in this paper
the Matsubara mechanism as a
path-integral formalism on $R^{D-2}\times S^{1}\times S^{1}$
to deal simultaneously with
 spatial constraints in a subspace of dimension $d=2$ (the transverse dimensions
of a wire having a rectangular cross-section).
These ideas have been
applied in different physical situations\cite{Ademir,GNN,JMario,kha1}.

We begin by briefly presenting the study of the U(1) Scalar
Electrodynamics in the transverse unitarity gauge, along the lines
developed in refs.\cite{isaque,camarda}, and afterwards presenting
a general procedure to treat a massive field theory in a
$D$-dimensional Euclidean space, compactified in a $d$-dimensional
subspace, with $d\leq D$. This permits to extend to an arbitrary
subspace some results in the literature for the behavior of field
theories in presence of spatial boundaries
\cite{JMario,Ademir,FAdolfo}. After describing the general
formalism, we focus on the particularly interesting case of $d=2$.
We investigate the behavior of the system taking into account
gauge fluctuations, which means that charged transitions are
included in our work. We are particularly interested in the
problem of how the critical behavior depends on the dimensions
$L_{i}$. This study is done by means of the Gaussian Effective
Potential (GEP) as developed in Refs. \cite{Gep1,Gep2,Gep3,Gep4},
together with the spatial compactification mechanism mentioned
above, introduced in  recent publications\cite{Ademir,JMario}.

\section{The Gaussian effective potential for the
Ginzburg-Landau model}

We start from the Hamiltonian density of
the GL model in Euclidean $d$-dimensional space (unless explicitly stated
we work in Natural Units (NU), $\hbar \,=\;c\,=\;k\,=\;1$)
written in the form \cite{Kleinert},
\begin{equation}
{\cal H'}= \frac{1}{4}F_{\mu\nu}F^{\mu\nu}+\frac{1}{2}|
(\partial_{\mu} -ieA_{\mu})\Psi |^{2}
+ \frac{1}{2}m_{0}^{2}|\Psi|^{2} +\lambda (|\Psi |^{2})^{2},
\label{GL}
\end{equation}
where $\Psi$ is a complex field, and $m_{0}$ is the bare mass.
The components of the transverse magnetic field,
$F_{\mu\nu}=\partial_{\mu}A_{\nu}-\partial_{\nu}A_{\mu}$
($\mu,\nu=1,...,d$) are related to the $d$-dimensional potential vector by the
well
known equation,

\begin{equation}
\frac{1}{2}F_{\mu\nu}F^{\mu\nu}=|{\bf \nabla}\times {\bf A}|^{2}.
\label{gau}
\end{equation}
In order to obtain only physical degrees of freedom, we can introduce two real
fields
instead of the complex field $\Psi$, assuming a transverse unitarity gauge.
We can define the field in terms of two real fields, as $\Psi=\phi
e^{i\gamma}$, together with the gauge transformation ${\bf A}\rightarrow
{\bf A}-1/e {\bf \nabla}\gamma$. The unitarity gauge makes
the original transverse field to acquire a longitudinal component
${\bf A}_{L}$ proportional to ${\bf \nabla}\gamma$. Then the original
functional integration over $\Psi$ and $\Psi^{*}$ in the generating
functional of correlation functions,  becomes an integration over
$\phi,\; {\bf A}_{T}$ and ${\bf A}_{L}$. The longitudinal component of
the vector potential can be integrated out, leading to
the generating functional (up to constant terms),

\begin{equation}
Z[j]=\int D\phi\;DA_{T}]exp \left[ -\int d^{d}x {\cal H} + \int
d^{d}x \;j\phi \right],
 \label{part}
\end{equation}
where the Hamiltonian is

\begin{equation}
{\cal H}=\frac{1}{2}(\nabla\phi)^2 + \frac{1}{2}m_{0}^{2}\phi^{2}
+\lambda\phi^{4}
+ \frac{1}{2}e^{2}\phi^{2}A^{2}+\frac{1}{2}({\bf
\nabla}\times {\bf A})^{2} + \frac{1}{2\epsilon}({\bf \nabla}
\cdot {\bf A})^{2}. \label{Lagrangeana}
\end{equation}
We have introduced above a gauge fixing term, the limit $\epsilon \rightarrow 0$
being taken later on after the calculations have been done.
In Eq.(\ref{Lagrangeana}) and in what follows, unless explicitly stated,
${\bf A}$ stands for the {\it transverse} gauge field.

The Gaussian effective potential can be obtained
from Eq.(\ref{Lagrangeana}), performing a
shift in the scalar field in the form $\phi = \tilde{\phi} + \varphi $,
which allows to write the Hamiltonian in the form
\be
{\cal H}={\cal H}_{0} + {\cal
H}_{int},
\label{H1}
\ee
 with ${\cal H}_{0}$ being the free part and ${\cal
H}_{int}$ the interaction part, given respectively by

\begin{equation}
{\cal H}_{0}= \left[ \frac{1}{2}(\nabla
\tilde{\phi})^2+\frac{1}{2} \Omega^{2}\tilde{\phi}^{2} \right]
+ \left[\frac{1}{2}({\bf \nabla}\times {\bf A})^{2} +\frac{1}{2}
\Delta^{2}A_{\mu}A^{\mu} +\frac{1}{2\epsilon}({\bf \nabla} \cdot
{\bf A})^{2}\right], \label{H0}
\end{equation}
and

\begin{equation}
{\cal H}_{int} = \sum_{n=0}^{4}
v_{n}\tilde{\phi}^{n}+\frac{1}{2}\left(
e^{2}\varphi^{2}-\Delta^{2} \right)A_{\mu}A^{\mu}
+ \frac{1}{2}e^{2}\tilde{\phi} A_{\mu}A^{\mu}\varphi+\frac{1}{2}
e^{2}A_{\mu}A^{\mu}\varphi^{2}, \label{Hint}
\end{equation}
where

\begin{eqnarray}
v_{0}&=&\frac{1}{2}m_{0}^{2}\varphi^{2}+\lambda\varphi^{4},\label{v0} \\
v_{1}&=&m_{0}^{2}\varphi+4\lambda\varphi^{3}, \label{v1} \\
v_{2}&=&\frac{1}{2}m_{0}^{2}\varphi^{2}+6\lambda\varphi^{2}
-\frac{1}{2}\Omega^{2},  \label{v2} \\
v_{3}&=&4\lambda\varphi,  \label{v3} \\
v_{4}&=&\lambda. \label{v4}
\end{eqnarray}
It is clear from Eqs. (\ref{H1},
(\ref{H0}) and (\ref{Hint}), that ${\cal H}$ describes two interacting fields, a
real scalar field $\phi$ of mass $\Omega$ and a real vector
gauge field {\bf A} of mass $\Delta$.

The effective potential, which is defined by
\begin{equation}
V_{eff}[\varphi]= \frac{1}{V}\left[ -lnZ + \int d^{d}x
j\varphi+\int d^{d}xJ_{\nu}A_{\nu} \right], \label{defpot}
\end{equation}
where $V$ is the total volume, can be obtained at first
order from standard methods from perturbation
theory. One can find, from Eqs. (\ref{part}),(\ref{H0}) and
(\ref{Hint}),

\begin{eqnarray}
V_{eff}[\varphi]&=& I_{1}^{d}(\Omega) + 2I_{1}^{d}(\Delta) +
\frac{1}{2}m_{0}^{2}\varphi^{2}+\lambda\varphi^{4}
 + \frac{1}{2} \left[ m_{0}^{2} - \Omega^{2}+
12\lambda\varphi^{2}+6\lambda I_{0}^{d}(\Omega) \right]
I_{0}^{d}(\Omega)\nonumber \\
&& + \left[ e^{2} \left( \varphi^{2} + I_{0}^{d}(\Omega)\right) -
\Delta^{2} \right]I_{0}^{d}(\Delta), \label{poteff}
\end{eqnarray}
where,

\be
I_{0}^{d}(M) = \int \frac{d^{d}k}{(2\pi)^{d}}\frac{1}
{k^{2}+M^{2}},
\label{I0}
\ee
and
\be
I_{1}^{d}(M) = \frac{1}{2} \int \frac{d^{d}k}{(2\pi)^{d}}
\ln(k^{2}+M^{2}),
\label{I1}
\ee
with $k=(k_{1},...,k_{d})$ being the $d$-dimensional momentum.

The Gaussian effective potential is derived by the requirement that
$V_{eff}[\varphi]$
must be stationary under variations of the masses $\Delta $ and
$\Omega $. This means that values $\overline{\Omega}$ and $\overline{\Delta}$
for the masses $\Omega$ and $\Delta$ should be found such that the conditions,

\begin{equation}
 \left(\frac{\partial V_{eff}}{\partial
\Omega^{2}}\right)_{\Omega^{2}=\overline{\Omega}^{2}}
 = 0, \label{sc1}
\end{equation}
\begin{equation}
 \left(\frac{\partial V_{eff}}{\partial
\Delta^{2}}\right)_{\Delta^{2}= \overline{\Delta}^{2}}= 0,
\label{sc2}
\end{equation}
 be simultaneously satisfied.
  These conditions generate the gap equations,

\begin{eqnarray}
\overline{\Omega}&=& m_{0}^{2} + 12\lambda \varphi^{2} +12\lambda
I_{0}^{d}(\overline{\Omega}) + 2e^{2} I_{0}^{d}
(\overline{\Delta}), \label{gap1} \\
\overline{\Delta}&=& e^{2} \varphi^{2}+ e^{2}
I_{0}^{d}(\overline{\Omega}). \label{gap2}
\end{eqnarray}
Replacing  $\Omega$ and $\Delta$ in Eq.(\ref{poteff})
by the solutions $\overline{\Omega}$ and $\overline{\Delta}$, of
Eqs.(\ref{gap1}) and
(\ref{gap2}) we obtain for the GEP the formal expression,

\begin{equation}
\overline{V}_{eff}[\varphi] = I_{1}^{d}(\overline{\Omega}) +
2I_{1}^{d}(\overline{\Delta})+\frac{1}{2} m_{0}^{2}\varphi^{2}
+\lambda\varphi^{4}
- 3\lambda [I_{0}^{d}(\overline{\Omega})]^{2}-e^{2}
I_{0}^{d}(\overline{\Omega})I_{0}^{d}(\overline{\Delta}).
\label{GEP}
\end{equation}
Notice that Eqs.(\ref{gap1}) and (\ref{gap2}) are a pair of very involved
coupled
equations, and no analytical solution for
them has been found, they can be solved only by numerical methods.
Later on we will see that this difficulty, in the limit of criticality can be
surmounted.

Next we intend to write an expression for the Gaussian
mass, $\overline{m}$, obtained in our case from the standard
prescription, as the second derivative of the {\it Gaussian} effective potential
for $\varphi=0$. To calculate the second derivative
of $\overline{V}_{eff}$ with respect to $\varphi$, we remark from
Eqs.(\ref{gap1}) and (\ref{gap2}) that $\overline{\Omega}^{2}$ and
$\overline{\Delta}^{2}$ also depend on $\varphi$ according to the
relations

\begin{eqnarray}
\frac{d\overline{\Omega}^{2}}{d\varphi} &=& \frac{24\lambda\varphi
-e^{2}I_{-1}^{d}(\overline{\Delta})\frac{d\overline{\Delta}^{2}}
{d\varphi}}{1+6\lambda I_{-1}^{d}(\overline{\Omega})}, \label{rel1}\\
\frac{d\overline{\Delta}^{2}}{d\varphi} &=& 2e^{2}\varphi
-\frac{1}{2}e^{2}I_{-1}^{d}(\overline{\Delta})
\frac{d\overline{\Omega}^{2}}{d\varphi}, \label{rel2}
\end{eqnarray}
where

\begin{equation}
I_{-1}^{d}(M)=2 \int \frac{d^{d}k}{(2\pi)^{d}}
\frac{1}{(k^{2}+M^{2})^{2}}. \label{I-1}
\end{equation}
Replacing Eq.(\ref{rel2}) in (\ref{rel1}) we get,

\begin{equation}
\frac{d\overline{\Omega}^{2}}{d\varphi} = \frac{\left[ 24\lambda
-2e^{4}I_{-1}^{d}(\overline{\Delta})\right] \varphi} {1+\left[
6\lambda - \frac{1}{2}e^{4}I_{-1}^{d}(\overline{\Omega}) \right]
I_{-1}^{d}(\overline{\Omega})}, \label{rel3}
\end{equation}
and the second derivative of the GEP with respect to $\varphi$ is given by,
\begin{eqnarray}
\frac{d^{2} \overline{V}_{eff}}{d \varphi^{2}}&=& m_{0}^{2}+12\lambda
\varphi^{2}
+ 12\lambda I_{0}^{d}(\overline{\Omega})+2e^{2}I_{0}^{d}
(\overline{\Delta})+2 e^{4} \varphi^{2}I_{-1}^{d}(\overline{\Delta})
\nonumber \\
&-&\frac{\left[ 6\lambda + \frac{1}{2}
e^{4}I_{-1}^{d}(\overline{\Delta}) \right]\left[
24\lambda-2e^{4}I_{-1}^{d}(\overline{\Delta})\right]
\varphi^{2}}{1+\left[ 6\lambda - \frac{1}{2}e^{4}I_{-1}
(\overline{\Omega}) \right]I_{-1}^{d}(\overline{\Omega})}. \nonumber\\
\label{2derV}
\end{eqnarray}
Thus we have the formula for the Gaussian mass,

\begin{eqnarray}
\overline{m}^{2} & \equiv & \left. \frac{d^{2} V_{eff}}{d
\varphi^{2}}\right|_{\varphi=0}=  m_{0}^{2}+ 12\lambda
I_{0}^{d}(\overline{\Omega}_{0})
+2e^{2}I_{0}^{d}(\overline{\Delta}_{0}),  \label{grm}
\end{eqnarray}
where $\overline{\Omega}_{0}$ and $\overline{\Delta}_{0}$ are respectively
solutions
for $\overline{\Omega}$ and $\overline{\Delta}$ at $\varphi=0$, explicitly,

\begin{eqnarray}
\overline{\Omega}_{0}^{2}&=& m_{0}^{2} +12\lambda I_{0}^{d}
(\overline{\Omega}_{0})
+ 2e^{2} I_{0}^{d}(\overline{\Delta}_{0}),   \label{gap3} \\
\overline{\Delta_{0}}&=& e^{2}I_{0}^{d}(\overline{\Omega}_{0}).
\label{gap4}
\end{eqnarray}
Therefore, from Eqs.(\ref{grm}) and (\ref{gap3}) we get simply,
\begin{equation}
\overline{m}^{2}=\overline{\Omega}_{0}^{2}.  \label{mOm}
\end{equation}
Hence, we see from the gap equation
(\ref{grm}) that the Gaussian  mass obeys a generalized "Gaussian"
Dyson-Schwinger equation,
\begin{equation}
\overline{m}^{2}=m_{0}^{2}+ 12\lambda I_{0}^{d}(\overline{m})
+2e^{2}I_{0}^{d}\left(\sqrt{e^{2}I_{0}^{d}(\overline{m})}\right).
\label{gap5}
\end{equation}
This expression will be used later to describe the system in the neighbourhood
of
criticality.

\section{Critical behavior of the Ginzburg-Landau model compactified on two
spatial dimensions}

We will work in the approximation of
neglecting boundary corrections to the coupling constant. A
precise definition of the boundary-modified mass parameter will be
given later for the situation of $D=3$ with $d=2$,
corresponding  to a wire of
rectangular section $L_{1}\times L_{2}$.
We also consider  the limiting situation of
one of the transverse dimensions, $L_{2}\rightarrow \, \infty$,
which corresponds  to a film of thickness $L_{1}$.
We use Cartesian coordinates ${\bf r}=(x_{1},...,x_{d},{\bf z})$,
where ${\bf z}$ is a $(D-d)$-dimensional vector, with
corresponding momentum ${\bf k}=(k_{1},...,k_{d}, {\bf q})$, ${\bf
q}$ being a $(D-d)$-dimensional vector in momentum space. Then the
generating functional of correlation functions has the form,
\begin{equation}
{\cal Z}=\int {\cal D}\varphi ^{\dagger }{\cal D}\varphi \exp \left(
-\int_{0}^{{\bf L}}d^{d}r\int d^{D-d}{\bf z}\;{\cal H}(\varphi,\nabla
\varphi)
\right) \,,  \label{part}
\end{equation}
where ${\bf L}=(L_{1},...,L_{d})$, and we are allowed to introduce
a generalized Matsubara prescription, performing the following
multiple replacements (compactification of a $d$-dimensional
subspace),
\begin{equation}
\int \frac{dk_{i}}{2\pi }\rightarrow \frac{1}{L_{i}}\sum_{n_{i}=-\infty
}^{+\infty }\;;\;\;\;\;\;\;k_{i}\rightarrow \frac{2n_{i}\pi }{L_{i}}
\;,\;\;i=1,2,...,d.
\label{Matsubara1}
\end{equation}
A simpler situation is the system confined simultaneously between
two parallel planes a distance $L_1$ apart from one another normal
to the $x_1$-axis and two other parallel planes, normal to the
$x_2$-axis separated by a distance $L_2$ (a ``wire" of rectangular
section). We emphasize however, that here we are considering an
Euclidean field theory in $D$ {\it purely} spatial dimensions, we
are {\it not} working in the framework of finite temperature field
theory. Temperature is introduced in the mass term of the
Hamiltonian by means of the usual Ginzburg-Landau prescription.

For our proposes we only need the calculation of the integral
given in equation (\ref{I0}) in the situation of confinement of
the present section. With the prescription (\ref{Matsubara1}), the
expression  corresponding to  equation (\ref{I0}) for the confined
system can be written in the form,

\begin{equation}
I_{0}^{D}(M)=\frac{1}{4\pi^2
L_{1}L_{2}..L_{d}}\sum_{n_{1},..,n_{d}=-\infty }^{\infty }\int
\frac{d^{D-d}q}{(a_{1}^{2}n_{1}^{2}
+a_{2}^{2}n_{2}^{2}+..+a_{d}^2n_{d}^2+c^{2}+ q^2)} \;,
\label{integral1}
\end{equation}

In the following, to deal with dimensionless quantities
in the regularization procedures, we introduce the parameters

\begin{equation}
c=\frac{m}{2\pi},\;\;a_{i}=\frac{1}{L_{i}},\;\;q_{i}=\frac{k_{i}}{2\pi}
\label{semdim}
\end{equation}

and using a well-known dimensional regularization formula
\cite{Ramond} to perform the integration over the ($D-d$)
non-compactified momentum variables, we have,
\begin{equation}
\\
\int \frac{d^{D-d}q}{(a_{1}^{2}n_{1}^{2}
+a_{2}^{2}n_{2}^{2}+..+a_{d}^2n_{d}^2+c^{2}+q^2)^s}=
\frac{\pi^{\frac{D-d}{2}}\Gamma(s-\frac{D-d}{2})}
{\Gamma(s)}\frac{1}{(a_{1}n_{1}^2+a_{2}n_{2}^2+..+a_{d}^2n_{d}^2+c^2)^{s-\frac{D-d}{2}}}
\,,
\label{potefet2}
\end{equation}
with which Eq.(\ref{integral1}) is written,
\begin{equation}
I_{0}^D(M)=\frac{\pi^{\frac{D-d}{2}}\Gamma(s-\frac{D-d}{2})}{4\pi^2
L_{1}L_{2}...L_{d}}\sum_{n_{1},..n_{d}=-\infty
}^{\infty}\frac{1}{(a_{1}n_{1}^2+a_{2}n_{2}^2+..+a_{d}^2n_{d}^2+c^2)^{s-\frac{D-d}{2}}}.
\label{formula}
\end{equation}
The sum in Eq.(\ref{formula}) is one of the Epstein-Hurwitz zeta-functions,
defined by,
\begin{eqnarray}
A_{d}^{c^{2}}(\nu ;a_{1},...,a_{d})& = & \sum_{n_{1},...,n_{d} =
-\infty} ^{\infty }(a_{1}^{2}n_{1}^{2}+\cdots
+a_{d}^{2}n_{d}^{2}+c^{2})^{-\nu }
\label{zeta}
\end{eqnarray}\\
where $\nu=s-\frac{D-d}{2}$. Next we can proceed generalizing to
several dimensions the mode-sum regularization prescription
described in Ref. \cite{Elizalde}. This generalization has been
done in \cite{Ademir} and we briefly describe here its main steps.
>From the identity,
\begin{equation}
\frac{1}{\Delta ^{\nu }}=\frac{1}{\Gamma (\nu )}\int_{0}^{\infty
}dt\;t^{\nu -1}e^{-\Delta t},
\end{equation}
and using the representation for Bessel functions of the
third kind, $K_{\nu}$,
\begin{equation}
2(a/b)^{\frac{\nu }{2}}K_{\nu }(2\sqrt{ab})=\int_{0}^{\infty
}dx\;x^{\nu -1}e^{-(a/x)-bx},  \label{K}
\end{equation}
we obtain after some rather long but straightforward manipulations
\cite{Ademir},
\begin{eqnarray}
A_{d}^{c^{2}}(\nu ;a_{1},...,a_{d}) &=&\frac{2^{\nu
-\frac{d}{2}+1}\pi ^{2\nu -\frac{d}{2}}}{a_{1}\cdots a_{d}\,\Gamma
(\nu )}\left[ 2^{\nu - \frac{d}{2}-1}\Gamma \left(\nu
-\frac{d}{2}\right) \left( 2 \pi c \right)^{d-2\nu } +
2\sum_{i=1}^{d}\sum_{n_{i}=1}^{\infty }\left(\frac{n_{i}} {2\pi c
a_{i}}\right)^{\nu - \frac{d}{2}}K_{\nu
-\frac{d}{2}}\left( \frac{2\pi c n_{i}}{a_i} \right)\right.  \nonumber \\
&&\left.+\cdots +2^{d}\sum_{n_{1},...,n_{d}=1}^{\infty }
\left(\frac{1}{2\pi c} \sqrt{\frac{n_{1}^{2}}{a_{1}^{2}} + \cdots
+ \frac{n_{d}^{2}}{a_{d}^{2}}} \right)^{\nu - \frac{d}{2}} K_{\nu
- \frac{d}{2}} \left( 2\pi c \sqrt{\frac{n_{1}^{2}}{a_{1}^{2}} +
\cdots + \frac{n_{d}^{2}}{a_{d}^{2}}} \right)\right] .
\label{zeta4}
\end{eqnarray}

In our case $s=1$ and for wires $d=2$, and Eq.(\ref{zeta4})
becomes,
\begin{eqnarray}
A_{2}^{c^2}(2-\frac{D}{2};L_{1},L_{2})&=&~\frac{L_{1}L_{2}2^{2-\frac{D}{2}}\pi^{3-D}}
{\Gamma(2-\frac{D}{2})}~
\left[~2^{-\frac{D}{2}}\Gamma(1-\frac{D}{2})M^{D-2}+\right.\nonumber\\
&&\left.+2\sum_{n_{1}=1}^{\infty}\left
(\frac{L_{1}^2n_{1}}{M}\right)^{1-\frac{D}{2}}K_{1-\frac{D}{2}}(ML_{1}n_{1})+
2\sum_{n_{2}=1}^{\infty}\left(\frac{L_{2}^2n_{2}}{M}\right)^{1-\frac{D}{2}}
K_{1-\frac{D}{2}}(ML_{2}n_{2})+\right.\nonumber\\
&&\left.+2^2\sum_{n_{1},n_{2}=1}^{\infty}\left(\frac{\sqrt
{L_{1}^2n_{1}^2+L_{2}^2n_{2}^2}}{M}
 \right)^{1-\frac{D}{2}}K_{1-\frac{D}{2}}\left(M\sqrt{L_{1}^2n_{1}^2+
 L_{2}^2n_{2}^2}\right)~\right].
 \label{formula2}
\end{eqnarray}
Thus, inserting the equation (\ref{formula2}) into equation
(\ref{formula}) and making some algebraic manipulations, we obtain

\begin{eqnarray}
I_{0}^D(M)&=&
2^{-D}\pi^{-\frac{D}{2}}\Gamma(1-\frac{D}{2})M^{D-2}+
2^{1-\frac{D}{2}}\pi^{-\frac{D}{2}}\left[\sum_{n_{1}=1}^{\infty}(\frac{L_{1}n_{1}}{M})^
{1-\frac{D}{2}}K_{1-\frac{D}{2}}(ML_{1}n_{1})\right. \nonumber\\
&&\left. +\sum_{n_{2}=1}^{\infty}(\frac{L_{2}n_{2}}{M})^
{1-\frac{D}{2}}K_{1-\frac{D}{2}}(ML_{2}n_{2})+2\sum_{n_{1},n_{2}=1}^{\infty}
(\frac{\sqrt{L_{1}^2n_{1}^2+L_{2}^2n_{2}^2}}{M})^{1-\frac{D}{2}}K_{1-\frac{D}{2}}
(M\sqrt{L_{1}^2n_{1}^2+L_{2}^2n_{2}^2})\right].
 \label{integral2}
\end{eqnarray}
where $K_{\nu}$ are the Bessel function of third kind.

\section{Critical Behavior for Wires}

In our case, we are
analyzing a superconducting type-I wire, which obeys  the following
conditions \cite{Kleinert},
\begin{equation}
\xi(T)=(\overline{m})^{-1}>>\lambda(T)>>L,
 \label{condition1}
 \end{equation}
 where  $L$ is the smallest one of the
linear dimensions in the rectangular cross-section area of the wire $A=L_{1} \times
L_{2}$,
and $\xi(T)$ and $\lambda(T)$ are respectively
the   critical correlation length and the critical penetration
depth defined by,
 \begin{equation}
 \xi(T)=\frac{\xi_{0}}{|t|^{1/2}}\;;\;\;\;\lambda(T)=\frac{\lambda_{0}}{|t|^{1/2}}\;;\;\;\;
 t=\frac{T-T_{c}}{T_{c}},
 \label{parametros}
 \end{equation}
where  $T_{c}$ is the transition temperature, $\xi_{0}$
 and $\lambda_{0}$ the intrinsic coherence length and London penetration depth,
 respectively.
In the case where one of the linear dimensions is much larger than
the other, $L_{1}\geq L_{2}$ or  $L_{2}\geq L_{1}$, we should
retrieve a film-like behavior,
 and L in Eq.(\ref{condition1}) would be the
thickness of the film.    After that, we can take $M=\overline{m}$ in
 Eq.(\ref{integral2}) and restrict ourselves to the
 neighborhood of criticality, that is to the region defined by
 $\overline{m}\approx 0$. Then we can use the
 asymptotic formula,
\begin{equation}
 K_{|\nu|}(z)=\frac{1}{2}\Gamma(|\nu|)(\frac{2}{z})^{|\nu|}\;;\;\;z\sim0
 \label{aproximation}
 \end{equation}
which allows, near criticality, to write equation (\ref{integral2}) in the form
\begin{equation}
I_{0}^{D}(\overline{m}\approx0)= 2^{-
\frac{D}{2}}\pi^{-\frac{D}{2}}\Gamma(\frac{D}{2}-1)\left[\sum_{n_{1}=1}^{\infty}
\frac{2^{\frac{D}{2}-1}}{(L_{1}n_{1})^{\frac{D}{2}-1}}+\sum_{n_{2}=1}^{\infty}
\frac{2^{\frac{D}{2}-1}}{(L_{2}n_{2})^{\frac{D}{2}-1}}+2\sum_{n_{1},n_{2}=1}^{\infty}
\frac{2^{\frac{D}{2}-1}}{(L_{1}^2n_{1}^2+L_{2}^2n_{2}^2)^{\frac{D}{2}-1}}\right]
\label{integral3}
\end{equation}
or,
\begin{equation}
I_{0}^{D}(\overline{m}\approx 0)\approx
\frac{\Gamma(\frac{D}{2}-1)}{2\pi^{\frac{D}{2}}}\left[\left(\frac{1}{L_{1}^{D-2}}+
\frac{1}{L_{2}^{D-2}}\right)\zeta(D-2)+2E_{2}(\frac{D-2}{2};L_{1},L_{2})\right]
\label{E2aproxim}
\end{equation}
where  $E_{2}\left(\frac{D-2}{2};L_{1},L_{2}\right)$ is the
generalized 2-dimensional Epstein $zeta$-function defined by\cite{Kirsten}
\begin{equation}
E_{2}\left(\frac{D-2}{2};L_{1},L_{2}\right) =
\sum_{n_{1},\,n_{2}=1}^{\infty} \left[ L_{1}
^{2}n_{1}^{2}+L_{2}^{2}n_{2}^{2}\right]^{-\left(
\frac{D-2}{2}\right)}\, .
\label{Z}
\end{equation}

In an analogous way as it is done for the Riemann $zeta$-function,
one can also
construct analytical continuations (and recurrence relations) for
the multidimensional Epstein-Hurwitz  $zeta$-functions which permit to write them
in terms of Kelvin and Riemann $zeta$ functions. To start one
considers the analytical continuation of the Epstein-Hurwitz
$zeta$-function given by \cite{Elizalde}
\begin{equation}
\sum_{n=1}^{\infty}\left( n^2 + p^2 \right)^{-\nu} = -\frac{1}{2}
p^{-2\nu} + \frac{\sqrt{\pi}}{2 p^{2\nu -1}\Gamma (\nu)} \left[
\Gamma\left( \nu-\frac{1}{2} \right) + 4\sum_{n=1}^{\infty} (\pi p
n)^{\nu-\frac{1}{2}} K_{\nu-\frac{1}{2}}(2\pi p n) \right] \; .
\label{EHfunc}
\end{equation}
Using this relation to perform one of the sums in (\ref{Z}) leads
immediately to the question of which sum is firstly evaluated.
Whatever the sum one chooses to
perform firstly, the manifest $L_{1} \leftrightarrow L_{2}$
symmetry of Eq.~(\ref{Z}) is lost; in order to preserve this
symmetry, we adopt here a symmetrized summation. Generalizing the
prescription introduced in \cite{Ademir}, we consider the
multidimensional Epstein function defined as the symmetrized
summation
\begin{equation}
E_{d}\left(\nu;L_{1},...,L_{d}\right) = \frac{1}{d !}
\sum_{\sigma} \sum_{n_1 = 1}^{\infty} \cdots \sum_{n_d =
1}^{\infty} \left[ \sigma_{1}^{2} n_{1}^{2} + \cdots +
\sigma_{d}^{2} n_{d}^{2} \right]^{\, - \nu} \; , \label{EfuncS}
\end{equation}
where $\sigma_{i}=\sigma(L_{i})$, with $\sigma$ running in the set
of all permutations of the parameters $L_1,...,L_d$, and the
summations over $n_1,...,n_d$ being taken in the given order.
Applying (\ref{EHfunc}) to perform the sum over $n_d$, one gets
\begin{eqnarray}
E_{d}\left(\nu;L_{1},...,L_{d} \right) & = & - \,\frac{1}{2\, d}
\sum_{i=1}^{d} E_{d-1}\left( \nu;...,{\widehat
{L_{i}}},...\right) \nonumber \\
 & & +\, \frac{\sqrt{\pi}}{2\, d\, \Gamma(\nu)} \Gamma\left( \nu - \frac{1}{2}
 \right) \sum_{i=1}^{d} \frac{1}{L_i} E_{d-1}
 \left( \nu-\frac{1}{2};...,{\widehat {L_{i}}},... \right)
 + \frac{2\sqrt{\pi}}{d\, \Gamma(\nu)} W_d \left( \nu-\frac{1}{2},L_1,...,L_d
 \right)\; , \label{Wd}
\end{eqnarray}
where the {\it hat} over the parameter $L_{i}$ in the functions
$E_{d-1}$ means that it is excluded from the set
$\{L_{1},...,L_{d}\}$ (the others being the $d-1$ parameters of
$E_{d-1}$), and
\begin{equation}
W_d\left(\eta;L_1,...,L_d\right) = \sum_{i=1}^{d} \frac{1}{L_i}
\sum_{n_1,...,n_d=1}^{\infty} \left( \frac{\pi n_i}{L_i \sqrt{(
\cdots + {\widehat {L_i^2 n_i^2}} + \cdots )}} \right)^{\eta}
K_{\eta}\left( \frac{2\pi n_i}{L_i} \sqrt{( \cdots + {\widehat
{L_i^2 n_i^2}} + \cdots )} \right) \; , \label{WWd}
\end{equation}
with $(\cdots + {\widehat {L_i^2 n_i^2}} + \cdots)$ representing
the sum $\sum_{j=1}^{d}L_{j}^{2}n_{j}^{2} \, - \,
L_{i}^{2}n_{i}^{2}$. In particular, noticing that $E_{1}\left(\nu;
L_j\right)=L_{j}^{-2\nu}\zeta(2\nu)$, one finds
\begin{eqnarray}
E_{2}\left(\frac{D-2}{2};L_{1} ^{2},L_{2}^{2}\right) & = &
-\frac{1}{4}\left(\frac{1}{L_{1}^{D-2}} +
\frac{1}{L_{2}^{D-2}}\right)
\zeta (D-2)  \nonumber \\
&&+\;\frac{\sqrt{\pi }\Gamma (\frac{D-3}{2})}{4\Gamma
(\frac{D-2}{2})}\left( \frac{1}{L_{1} L_{2}^{D-3}}+\frac{1}{L_{1}
^{D-3}L_{2}}\right) \zeta (D-3) +\frac{\sqrt{\pi }}{\Gamma
(\frac{D-2}{2})} W_{2}\left(\frac{D-3}{2};L_{1},L_{2}\right)\;.
\label{Z1}
\end{eqnarray}
Inserting  (\ref{Z1}) into (\ref{E2aproxim}) we get,

\begin{eqnarray}
I_{0}^{D}(\overline{m}\approx
0)&=&\frac{\Gamma(\frac{D-2}{2})}{4\pi^{\frac{D}{2}}}\left(\frac{1}{L_{1}^{D-2}}+
\frac{1}{L_{2}^{D-2}}\right)\zeta(D-2)+\frac{\Gamma(\frac{D-3}{2})}{4\pi^{D-1/2}}
\left(\frac{1}{L_{1}L_{2}^{D-3}}+
\frac{1}{L_{2}L_{1}^{D-3}}\right)\zeta(D-3) \nonumber\\
&& +\frac{W_{2}(\frac{D-3}{2};L_{1},L_{2})}{\pi^{D-1/2}},
\label{integral3}
\end{eqnarray}
where,
\begin{equation}
W_{2}(0;L_1,L_2) = \sum_{n_1,n_2 = 1}^{\infty} \left\{
\frac{1}{L_1} K_{0}\left( 2\pi\frac{L_{2}}{L_{1}}n_{1}n_{2}\right)
+ \frac{1}{L_2} K_{0}\left(
2\pi\frac{L_{1}}{L_{2}}n_{1}n_{2}\right) \right\}. \label{W2L}
\end{equation}
For $D=3$,
the first and second terms between the brackets in Eq.
(\ref{integral3}) are divergent due to the poles of the $\zeta$-function and
$\Gamma$-function, respectively. However, the whole expression in (\ref{integral3})
is {\it not} divergent, in fact
 these divergences cancel between themselves leaving a finite result
for $I_{0}^{3}(\overline{m}\approx 0)$. This can be easily seen if
we consider the following properties of the {\it zeta}-function,
\begin{equation}
\zeta(z)=\frac{1}{\Gamma(z/2)}\Gamma(\frac{1-z}{2})\pi^{z-\frac{1}{2}}\zeta(1-z)\,;\;\;\;\;\lim_{z\rightarrow 1}\left[\zeta(z)-\frac{1}{z-1}\right]=\gamma,
\label{reflection}
\end{equation}
where $\gamma\approx 0.5772$ is the Euler-Mascheroni constant.

The quantity $W_{2}(0;L_1,L_2)$, appearing in
Eq.(\ref{integral3}), involves complicated double sums; in particular it is very
difficult to be analytically handled for $L_1 \neq L_2$, what means that it is not
possible to take analytically limits such as $L_2 \rightarrow
\infty$ for finite $L_1$. Such a limit (which
corresponds to a film of thickness $L_1$) will be considered numerically.\\
Using Eq. (\ref{reflection})  in Eq. (\ref{integral3}) we obtain for $D=3$,

\begin{equation}
I_{0}^{3}(\overline{m}\approx0)\approx\frac{\gamma}{2\pi}\left(\frac{1}{L_{1}}+\frac{1}{L_{2}}
\right)+\frac{1}{\pi}W_{2}(0;L_{1}L_{2}),
 \label{integral5}
 \end{equation}
and
 Eq.(\ref{integral2}) becomes,

\begin{eqnarray}
I_{0}^{3}(\Delta_{0})&=&\frac{\Gamma(-\frac{1}{2})\Delta_{0}}{2^{3}\pi^{3/2}}+
\frac{1}{2^{1/2}\pi^{3/2}}\left[\sum_{n_{1}=1}^{\infty}\left(\frac{\Delta_{0}}{L_{1}n_{1}}\right)^{1/2}
K_{1/2}(\Delta_{0}L_{1}n_{1})+\sum_{n_{2}=1}^{\infty}\left(\frac{\Delta_{0}}{L_{2}n_{2}}\right)^{1/2}
K_{1/2}(\Delta_{0}L_{2}n_{2})+\right.\nonumber\\
&&\left.+\sum_{n_{1},n_{2}=1}^{\infty}
\left(\frac{\Delta_{0}}{\sqrt{L_{1}^2n_{1}^2+L_{2}^2n_{2}^2}}\right)^{1/2}K_{1/2}(\Delta_{0}
\sqrt{L_{1}^2n_{1}^2+L_{2}^2n_{2}^2})\right].
 \label{intD1}
\end{eqnarray}
Using the formula,
\begin{equation}
K_{1/2}(z)=\sqrt{\frac{\pi}{2z}}e^{-z} \; \label{approxK}
\end{equation}
and remembering the notation, $\Delta_{0}=\sqrt{e^2I_{0}^{D}(\overline{m}})$
we obtain, after some rearrangements,
\begin{eqnarray}
I_{0}^{3}(\sqrt{e^2I_{0}^{3}(\overline{m}}))&=&-\frac{e\sqrt{I_{0}^{3}(\overline{m}\approx
0)}} {4\pi}-\frac{ln(1-e^{-e
L_{1}\sqrt{I_{0}^{3}(\overline{m}\approx0)}})}{2\pi L_{1}}-
\frac{ln(1-e^{-eL_{2}\sqrt{I_{0}^{3}(\overline{m}\approx
0)}})}{2\pi L_{2}}+ \nonumber\\
&&+\frac{1}{\pi}\sum_{n_{1},n_{2}=1}^\infty
\frac{e^{-e\sqrt{I_{0}^{3}(\overline{m}\approx
0)}\sqrt{L_{1}^2n_{1}^2+L_{2}^2n_{2}^2}}}{\sqrt{L_{1}^2n_{1}^2+
L_{2}^2n_{2}^2}}.
\label{intD2}
\end{eqnarray}

Thus we can write the Gaussian gap equation (\ref{gap5})in the
neighbourhood of criticality in the form,

\begin{eqnarray}
\overline{m}^2&\approx& m_{0}^2+12\lambda
I_{0}^{3}(\overline{m}\approx0)-\frac{e^3}{2\pi}\sqrt{I_{0}^{3}(\overline{m}\approx0)}-
\frac{e^2}{\pi
L_{1}}ln(1-e^{-eL_{1}\sqrt{I_{0}^{3}(\overline{m}\approx0)}})+\nonumber\\
&&-\frac{e^2}{\pi
L_{2}}ln(1-e^{-eL_{2}\sqrt{I_{0}^{3}(\overline{m}\approx0)}})+\frac{2e^2}{\pi}
\sum_{n_{1},n_{2}=1}^\infty\frac{e^{-e\sqrt{I_{0}^{3}(\overline{m}\approx0)}
\sqrt{L_{1}^2n_{1}^2+L_{2}^2n_{2}^2}}}{\sqrt{L_{1}^2n_{1}^2+L_{2}^2n_{2}^2}}
\label{massa}
\end{eqnarray}
Taking $m_{0}^2=a(T/T_{0}-1)$, with $a>0$ we obtain from the above
equation for $\overline{m}^2=0$, the critical temperature as a
function of the wire transverse dimensions, $L_{1}\,\; L_{2}$ and the
bulk
transition temperature, $T_{0}$,

\begin{eqnarray}
T_{c}(L_{1},L_{2})&=&T_{0}\left[1-\frac{12\lambda}{a}I_{0}^{3}(\overline{m}\approx0)+\frac{e^3}{2a\pi}
\sqrt{I_{0}^{3}(\overline{m}\approx0)}+\frac{e^2}{a\pi
L_{1}}ln(1-e^{eL_{1}\sqrt{I_{0}^{3}(\overline{m}\approx0)}})+\right.\nonumber\\
&&\left. +\frac{e^2}{a\pi
L_{2}}ln(1-e^{eL_{2}\sqrt{I_{0}^{3}(\overline{m}\approx0)}})-\frac{2e^2}{a\pi}\sum_{n_{1},n_{2}=1}
^\infty\frac{e^{-e\sqrt{I_{0}^{3}(\overline{m}\approx0)}\sqrt{L_{1}^2n_{1}^2+L_{2}^2n_{2}^2}}}
{\sqrt{L_{1}^2n_{1}^2+L_{2}^2n_{2}^2}}\right].
\label{Tc}
\end{eqnarray}

For reasons that will become apparent later, let us
take $(L_{2}=p L_{1})$, $L_{1}\equiv L$, with $p$ a positive constant; then we get from
Eqs.(\ref{W2L}) and (\ref{integral5}),

\begin{equation}
I_{0}^{3}(\overline{m})=\frac{C_{p}}{ L}, \label{integralLp}
\end{equation}
where
\begin{equation}
C_{p}=\frac{\gamma}{2\pi}(\frac{p+1}{p})+\frac{1}{\pi}\left[\sum_{n_{1},n_{2}=1}^\infty(K_{0}(2\pi
pn_{1}n_{2})+\frac{1}{p}K_{0}(2\pi n_{1}n_{2}/p))\right].
\label{Cp}
\end{equation}
and the critical temperature (\ref{Tc}) is rewritten in the form,
 \begin{eqnarray}
 T_{c}(L,p)&=&T_{0}\left[1-\frac{12\lambda C_{p}}{a
 L}+\frac{e^3\sqrt{C_{p}}}{2\pi a\sqrt{L}}+\frac{e^2}{a\pi
 L}ln(1-e^{-e\sqrt{L C_{p}}})+\right. \nonumber\\
 &&\left.+\frac{e^2}{a\pi
 p L}ln(1-e^{-e p\sqrt{L C_{p}}})-\frac{2e^2}{a\pi
 L}\sum_{n_{1},n_{2}=1}^\infty \frac{e^{-e\sqrt{L
 C_{p}}\sqrt{n_{1}^2+p^2n_{2}^2}}}{\sqrt{n_{1}^2+p^2n_{2}^2}}\right].
 \label{TcLp}
 \end{eqnarray}

This equation describes the behavior of the critical temperature
of a type-I superconducting wire having a rectangular
cross-section, in terms of one of its transverse dimensions
($L_{1} \equiv L$) and of the parameter $p=L_2/L_1$. We recall
that in this expression all gauge fluctuations have been taken
into account by means of the Gaussian effective
potential. These contributions are given by the last four terms in Eq,(\ref{TcLp}).\\

\textsc{Comparison to some experimental results:}\\

Up to now we have used  Natural Units, $c=\hbar=k_{B}=1$. To
proceed, we restore  SI units, remembering that a factor
$1/k_{B}T_{0}$ is implicit in the exponent of equation
(\ref{part}). Next we rescale the fields and coordinates by
$\phi\longrightarrow\phi_{new}=\sqrt{\xi_{0}/k_{B}T_{0}}\phi $,
$A\longrightarrow A_{new}=\sqrt{\xi_{0}/k_{B}T_{0}}A $ and
$x\longrightarrow x_{new}=x/\xi_{0}$, with $\xi_{0}=0.18\hbar
v_{F}/k_{B}T_{0}$, being the intrinsic coherent length and $v_{F}$
the Fermi velocity of a particular material. This implies that, in
the Ginzburg-Landau model, the parameters $a$, $\lambda$  and  $e$
become dimensionless and are given by \cite{Kleinert},
\begin{equation}
a=1,~~ \lambda=111.08(\frac{T_{0}}{T_{F}})^2,~~
e\approx2.59\sqrt{\frac{\alpha v_{F}}{c}}, \label{parametros}
\end{equation}
where $T_{F}$ and $\alpha$ are respectively the Fermi temperature
and the fine structure constant. Thus, equation (\ref{TcLp}) is
rewritten as,

\begin{eqnarray}
~~~~T_{c}(L,pL)&=&T_{0}\left[1-1333.0(\frac{\xi_{0}}{L})(\frac{T_{0}}{T_{F}})^2C_{p}+
17.253x10^{-4}\sqrt{\frac{\xi_{0}}{L}}\sqrt{C_{p}v_{Fc}^3}+\right.\nonumber\\
&&\left.+15.594x10^{-3}(\frac{\xi_{0}}{L})v_{Fc}~ln(1-e^{-0.22128\sqrt{\frac{L}{\xi_{0}}}
\sqrt{C_{p}v_{Fc}}})+\right.\nonumber\\
&&\left.+15.594x10^{-3}(\frac{\xi_{0}}{pL})v_{Fc}~ln(1-e^{-0.22128p\sqrt{\frac{L}{\xi_{0}}}
\sqrt{C_{p}v_{Fc}}})+\right.\nonumber\\
&&\left.-0.031187(\frac{\xi_{0}}{L})v_{Fc}\sum_{n_{1}n_{2}=1}^\infty
\frac{e^{-0.22128\sqrt{\frac{L}{\xi_{0}}}\sqrt{C_{p}v_{Fc}}\sqrt{n_{1}^2+p^2n_{2}^2}}}
{\sqrt{n_{1}^2+p^2n_{2}^2}}\right]
\label{TcLexp}
\end{eqnarray}
As an application of this equation we plot  the behavior of
the critical temperature $T_{c}(L,p)$ as a function of  $1/L$  to
a superconducting wire  made from niobium, characterized by
$v_{F}=1.37\times 10^6$ m/s, $T_{0}=9.3$ K, and
$T_{F}=6.18\times 10^4$ K. We have considered three cases, with $p=1$,
$p=4$ and $p=10$  corresponding to  wires  of
different rectangular cross-sections.
\begin{figure}[c]
\epsfysize=5cm {\centerline{\epsfbox{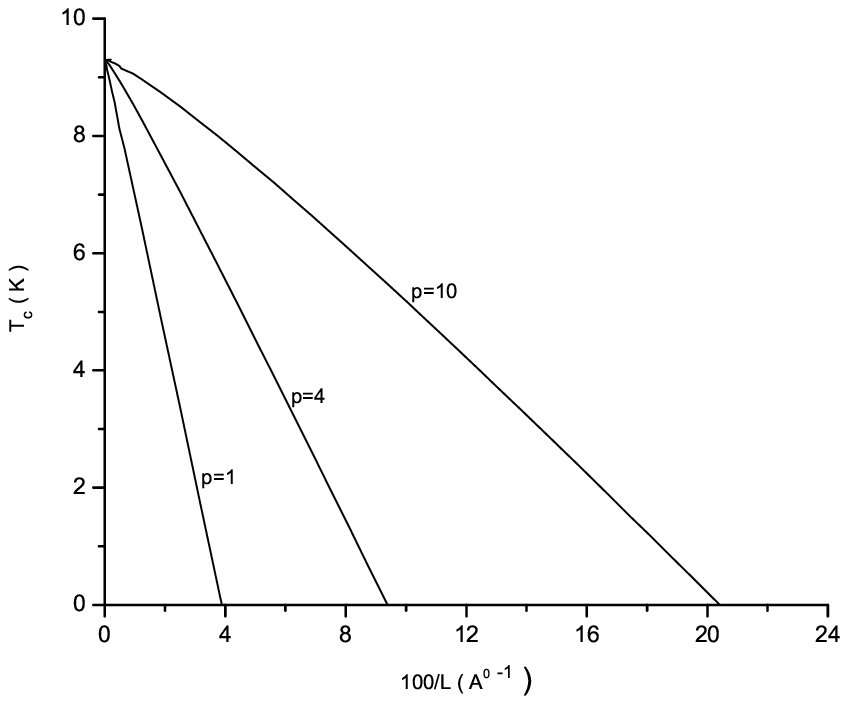}}}
\caption{Critical temperature $T_{c}$ as function of one of the
transverse dimensions $1/L$, from Eq.(\ref{TcLexp}) for a
superconducting wire made from Niobium
 of rectangular section cross $A=L\times pL$. The figure shows the
behavior of
 the critical temperature for the cases, $p=1$,
 $p=4$ and $p=10$, respectively.}
\end{figure}

In  $Fig.1$, we show the  quasi linear behavior of the critical
temperature of a niobium wire as a function of $L$, for $p=1$,
$p=4$ and $p=10$. These curves suggest the existence of minimal
values of $L$ (or of the rectangular wire cross-sections), for
which the surperconducting phase transition  is suppressed. The
approximate values for each case are: $L^{min}_{p=1}\approx 2.5
~nm$, $L^{min}_{p=4}\approx 1.0 ~nm$ and $L^{min}_{p=10}\approx
0.5 ~nm$ respectively.
\begin{figure}[c]
\epsfysize=5cm {\centerline{\epsfbox{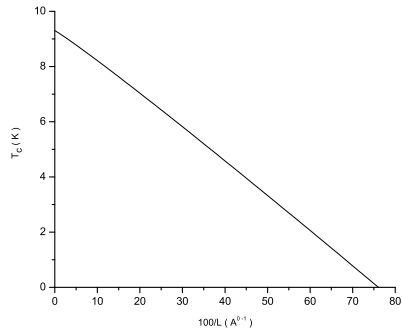}}}
\caption{Critical temperature $T_{c}$ as function of the inverse
thickness $1/L$, for a Niobium film}
\end{figure}
Also, it can be shown numerically that as we take
larger and larger values of $p$ ($p\rightarrow\infty$) in Eq.(\ref{TcLexp})
all the curves tend to coincide,
corresponding to a sample in the form of a film of thickness $L$.
A numerical analysis of Eq.(\ref{Cp}) shows that as $p\rightarrow\infty$,
the quantity $C_{p}$ tends to a constant value, $C_{p}=\frac{\gamma}{2\pi}$.
Replacing this value in Eq.(\ref{TcLexp}) we get the critical temperature for a film.
In Fig.2  we plot Eq.(\ref{TcLexp}) in this limiting situation,
which corresponds to a superconducting film of thickness $L$. In
this case we find a minimal film thickness  $L_{film}^{min}\approx 0.13 nm $.

\section{Conclusions}
In this paper we have considered the confined Ginzburg-Landau model,
  in the transverse unitarity
gauge, as a model to describe samples of superconducting materials
not in bulk form. To generate
the contributions from gauge fluctuations, we have used the
Gaussian effective potential \cite{Gep1,Gep2,Gep3,Gep4}, which allows to obtain
a gap
equation that can be treated with the method of recent
developments \cite{Ademir,JMario}. We have derived a critical
equation that describes the changes in the critical temperature
 due to confinement. Contributions from the
self interaction of the scalar field and from the gauge field
fluctuations can be clearly identified. Our approach suggests a minimal
size sustaining the existence of charged and non-charged
 transitions for both superconducting wires and films.
Such a kind of result could be of
practical interest to define  limits of miniaturization of
superconducting wires in  manufacturing  electronic circuits.
 It is interesting to compare
the minimal allowed value of $L\approx 2.5\,nm$ for a wire having a square cross-section
that we have obtained,
to the experimentally observed superconducting state in nanowires having
a cross-section diameter of $\approx 10\;nm$ \cite{nature}

Also, the behavior described in Eq.(\ref{TcLexp}) for
$p\rightarrow \infty$ (a film) could be contrasted with the linear
decreasing of the critical temperature with the inverse of the
film thickness, that has been found {\it experimentally} in
materials containing transition metals, for example, in PB
\cite{Strongin}, in W-Re alloys \cite{W-Re}, in Nb
\cite{Kodama,Park,Quateman}, Mo-Ge \cite{Graybeal}; for some of
these cases, this behavior has been explained in terms of
proximity, localization and Coulomb-interaction effects.
 We can clearly see from Eq.(\ref{TcLexp}),
as noticed above, that a linear decreasing of the critical
temperature with the inverse of film thickness is recovered when
we take $e=0$. Even though, with $e\neq 0$ a quasi-linear
decreasing of $T_{c}$ with $1/L$ can be directly obtained from
Eq.(\ref{TcLexp}), since the terms coming from effects of gauge
fluctuations are very small as compared to the term generated from
the self coupling.  Also a comparison may be done with recent
theoretical results for type II superconductors\cite{Luciano},
where a similar behavior of the critical temperature with the film
thickness has been found for non-charged transitions.

However, it should be noticed that we have used 3-dimensional values for
the phenomenological parameters. This means that we should restrict ourselves
only to relatively thin wires and films, which could be considered as essentially
3-dimensional objects;
 very thin wires and films {\it can not} be
physically accommodated in the context of our model. For instance,
the value we have obtained for $L_{film}^{min}(Nb) (\approx 0.13
\; nm)$ is of the order of magnitude of a few Bohr radii and
should be considered to correspond to a 2-dimensional system. So,
on physical grounds, it is beyond the domain of validity of the
model. Nevertheless our results obtained from the "pure" GL model
are in {\it qualitative} agreement with the experimentally
observed behaviors mentioned above. It is worth to emphasize that
the quasi-linear character of the decreasing of the  transition
temperature obtained in this paper, emerge solely as a topological
effect of the spatial compactification of the Ginzburg-Landau
model.

Also, it should be remarked that our formalism  can not account
for microscopic details present in real samples of materials nor
for the influence of manufacturing aspects, like the kind  of the
subtract on which a film is deposited, or the presence of
impurities.  This means that our expression for the transition
temperature should be further improved (in particular the GL
parameters), and more work is needed to better understand the
behavior of real samples. This will be the subject of future
investigation.

\begin{center}
{\bf Acknowledgements}
\end{center}
This work has been partially supported by the Brazilian agency CNPq (Brazilian
National Research Council). One of us, IR, also thank Pronex/MCT for
partial support.

\end{document}